# X-ray photoelectron spectroscopy, Raman and photoluminescence studies on formation of defects in Cu:ZnO thin films and its role in non-linear optical features

Albin Antony[1], P. Poornesh[1,*], I.V.Kityk[2], K.Ozga[2], J.Jedryka[2], P.Rakus[2] and A.Wojciechowski[2]

[1]Department of Physics, Manipal Institute of Technology, Manipal Academy of Higher Education, Manipal, Karnataka 576104, India
[2]Faculty of Electrical Engineering, Institute of Optoelectronics and Measuring Systems, Czestochowa University of Technology, Armii Krajowej 17, PL-42-201 Czestochowa, Poland

**Abstract.** The structural, optical, morphological and nonlinear optical properties of Cu:ZnO spray-coated films are studied. The surface morphology of Cu:ZnO thin films turned out to be homogenous, crack free and well covered with pea-shaped grains. The peak shift observed in the x-ray photoelectron spectroscopy spectra of the Cu:ZnO thin films infers the defect states present in the films. The satellite peak observed at 939.9 eV for Cu2P core-level spectra confirms the +2 oxidation state of Cu in the films. The formation of additional defect levels in the nanostruc-tures upon Cu doping was investigated using photoluminescence (PL) and Raman spectroscopy studies. The luminescent centers in the violet, blue and green spectral region were observed. The most prominent emission was centered at the blue color center for 5% Cu:ZnO thin films. The enhancement in the PL emission intensity confirms the increase in the defect state density upon Cu doping. The shifting of the UV emission peak to the visible region validates the increase in the non-radiative recombination process in the films upon doping. The phonon modes observed in Raman analysis around 439, 333 and 558 cm$^{-1}$ confirm the improvement in the crystallinity and formation of defect states in the films. X-ray diffraction reveals that the deposited films are of single-phase wurtzite ZnO structure with preferential growth orientation parallel (0 0 2) to the *C*-axis. The third-order optical susceptibility $x(3)$ has been increased from $3.5 \times 10^{-4}$ to $2.77 \times 10^{-3}$ esu due to the enhancement of electronic transition to different defect levels formed in the films and through local heating effects arising due to continuous wave laser illumination. The enhanced third harmonic generation signal investigated using a Nd:YAG laser at 1064 nm and 8 ns pulse width shows the credibility of Cu:ZnO films in frequency tripling applications.

*E-mail: poorneshp@gmail.com

## 1. Introduction

Recently, transparent conductive oxide materials based on Zinc oxide (ZnO) have gained intense research interest owing to their potential applications for optoelectronics and nonlinear devices [1]. The coexistence of excellent physical and chemical properties with a band gap of 3.3 eV, large exciton binding energy of 60 meV, its abundance and eco-friendly properties has made ZnO one of the most credible and widely studied materials to date [2-12]. The inherent properties of the ZnO material can be tuned by appropriate dopant material towards the preferred applications. The doping of transition metal ions into the ZnO lattice will lead to defect engineering and tailoring of certain properties of ZnO towards various practical applications [13-15]. Among the various transition metals, copper (Cu) ions can provide various interesting properties in ZnO thin films. The diffusion of the Cu atom into the ZnO lattice will replace either the Zn atom or interstitial Zn atoms forming structural defects. This can considerably affect the optical, structural, morphological, electrical and nonlinear optical properties of ZnO [13, 16-18].

The reports available on the doping, synthesis and study of Cu-doped ZnO nanostructures convey information about an emerging class of multifunctional Cu-doped ZnO devices that possess improved physical and chemical properties. Several deposition techniques for Cu-doped ZnO thin films have been reported, which include molecular beam epitaxy, RF sputtering, thermal evaporation, spray pyrolysis, hydrothermal method, sol-gel, etc [13, 18, 19]. As far as the practical applications are concerned, chemical methods are advantageous due to their flexibility of operation, simplicity, cost effectiveness and effective control of the doping concentration. The ability of large-area deposition with perfect stoichiometry and uniformity in the grown films makes spray pyrolysis more preferable than the other chemical methods [20].

Material with high nonlinear optical susceptibility is required to meet the present demand of nonlinear device applications. Although ZnO is investigated for optical nonlinearity, there is a lack of investigation into nonlinear mechanisms and the affecting parameters, which are essential for nonlinear optical devices. The studies related to defect states in the Cu:ZnO thin films grown by spray pyrolysis as a possible candidate for nonlinear device applications is limited and requires additional investigation. In-depth study of the enhancement in the intermediate defect states and other physical properties of Cu doping will be a possible pathway to understand the origin of the appearance of optical nonlinearity and to design the intended versatile nonlinear optical devices. The current work aims at understanding the basic physical mechanisms of optical nonlinearity in Cu:ZnO thin films and the parameters which affect the nonlinearity that has been investigated. Apart from this, in this report we explore the influence of defect states and other physical properties of the third harmonic generation (THG) process for Cu:ZnO thin films using UV-vis spectroscopy, x-ray diffraction (XRD), atomic force microscopy (AFM), x-ray photoelectron spectroscopy (XPS), photoluminescence (PL) spectroscopy and Raman spectroscopy analysis.

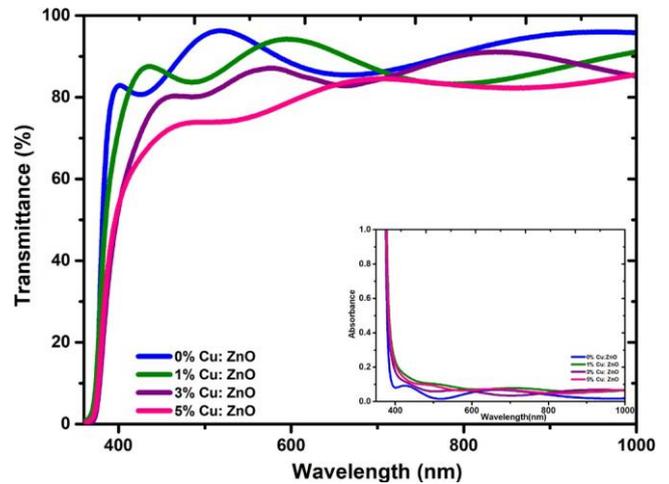

**Figure 1.** Optical transmittance and absorbance spectra of Cu:ZnO thin films.

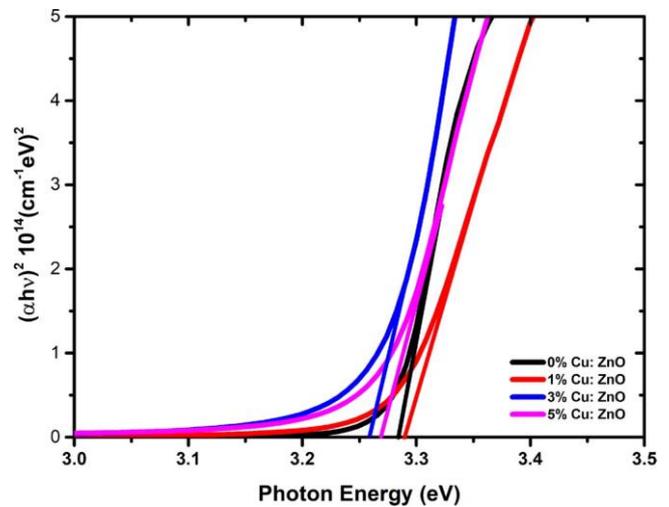

**Figure 2.** Tauc plot of Cu:ZnO thin films at different doping concentration.

## 2. Experimental

Pure and Cu-doped ZnO thin films were grown by spray pyrolysis deposition technique with copper concentration varying from 0 wt.%-5 wt.%. The precursor solution used for the deposition consists of zinc chloride ($ZnCl_2$) and copper(II) chloride dihydrate ($CuCl_2 \cdot 2H_2O$) dissolved in double-distilled water. The molarity of the solutions was maintained at 0.05 M. The glass substrate used for the deposition was ultrasonically cleaned with double-distilled water, acetone and 2-propanol before the deposition. The deposition time and other spray parameters were kept fixed in order to maintain a uniform thickness (~310 nm) of the studied films. The experimental procedures and spray parameters followed for the deposition of Cu:ZnO thin films is reported elsewhere [21].

The crystal structure and strain in the films was characterized by glancing angle XRD (Rigaku SmartLab, Cu K$a$ $\lambda = 1.54$ Å) with a step size of 0.020 and grazing angle of 0.60. The composition and oxidation states of the elements present in the films was determined by XPS using an Al K$a$ x-ray source of 1486.6 with 20 eV pass energy. The energy band gap and linear optical properties were studied using a UV-vis spectrophotometer (Shimadzu-1800) with a spectral resolution of 1 nm. An AFM (Bruker Icon) in tapping mode configuration was used to display the surface morphology of the films. The radiative and non-radiative transitions in the Cu:ZnO thin films were studied using PL (Horiba fluromax-4, $\lambda$excitation = 325 nm) and Raman (Horiba JOBIN YVON LabRAM HR, $\lambda$excitation = 532 nm) spectroscopy analysis. Z-scan technique was employed to determine the nonlinear optical properties of the films. Third harmonic ($3\omega$) frequency conversion efficiency of the films was carried out using a fundamental beam generation of 8 ns pulse from a Nd:YAG laser with a wavelength of 1064 nm and frequency repetition rate of 10 Hz. The THG signals were monitored before and after laser treatment using a coherent continuous wave (CW) laser light at a wavelength of 532 nm and power of about 200 mW with respect to $3\omega$ generation at 355 nm for a beam diameter varying within 4-10 mm. The CW illumination was performed for a short time in order to avoid sample overheating.

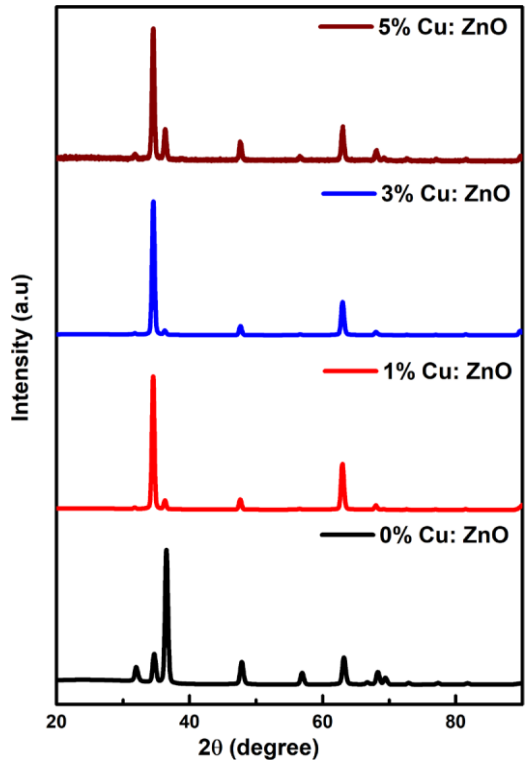

**Figure 3.** XRD pattern of Cu:ZnO thin films.

**Table 1.** Structural parameters of Cu:ZnO thin films.

| Conc. (wt %) | Crystalline size ($D$) nm | Dislocation density $\delta$ ($10^{15}$ lines m$^{-2}$) | Strain $\epsilon$ ($10^{-3}$) |
| --- | --- | --- | --- |
| 0 | 15.6 | 4.08 | 2.21 |
| 1 | 18.35 | 2.96 | 1.88 |
| 3 | 17.76 | 3.17 | 1.95 |
| 5 | 18.52 | 2.91 | 1.87 |

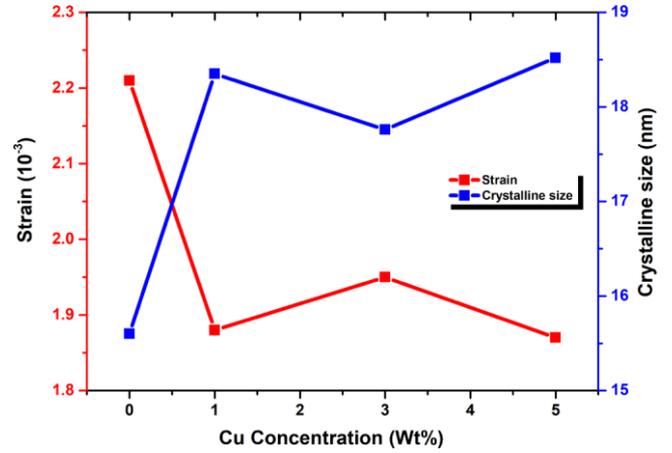

**Figure 4.** Variations of crystalline size and strain with Cu concentration.

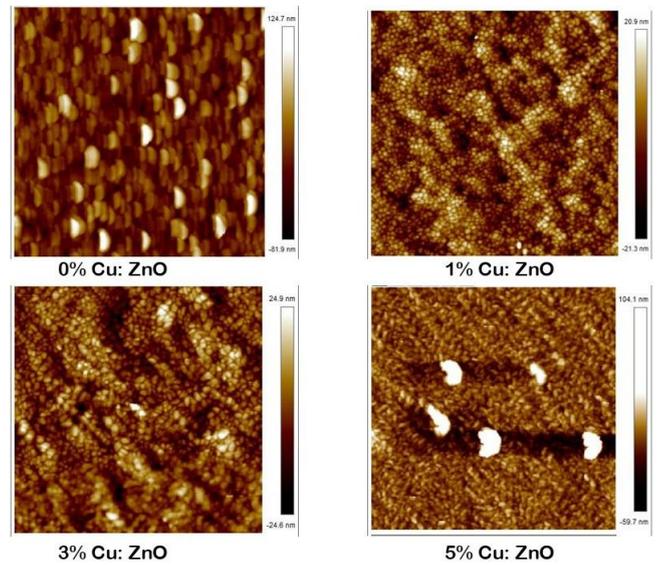

**Figure 5.** 2D AFM images of Cu:ZnO thin films.

## 3. Results and discussions

### 3.1. Optical, structural and morphological properties

The optical transmittance and absorbance spectra of undoped and 1%, 3% and 5% Cu-doped ZnO thin films in the 350-1000 nm spectral range was corrected for substrate effects and is depicted in figure 1. An increase of Cu doping concentration diminished the transmittance of the films. The undoped film shows an average transmittance of 85%, which decreases to 75% upon Cu incorporation. The decrement in the transmittance observed may be associated with the scattering of incident photons by the Cu ion absorption incorporated in the host ZnO lattice. On the other hand, the enhancement in the metal-to-oxygen ratio also led to the decrease in the transmittance of the Cu: ZnO thin films [22]. The red spectral shift observed in the absorption edge shown in the inset of figure 1 confirms the shrinkage of the optical band gap of the film upon Cu doping. The spectral shift in the absorption edge is likely due to the formation of Zn and O vacancies, which reduce the carrier concentration of the film upon Cu doping. From the recorded absorption spectra, the energy band gap of the films was found by using the Tauc plot method [23], as shown in figure 2. The bandgap values of the deposited Cu:ZnO thin films have been varied within an energy range of 3.29–3.26 eV. The observed reduction in the band gap of the films is attributed to the incompatibility in the electronegativity size of Cu and Zn atoms in the ZnO lattice, which in turn give rise to many chemical and magnetic effects [24, 25]. Apart from this, the reduction in the number of charge carriers donated by the defect states present in the films can be a possible reason for the slight reduction in the bandgap energy of the Cu:ZnO nanostructures [24].

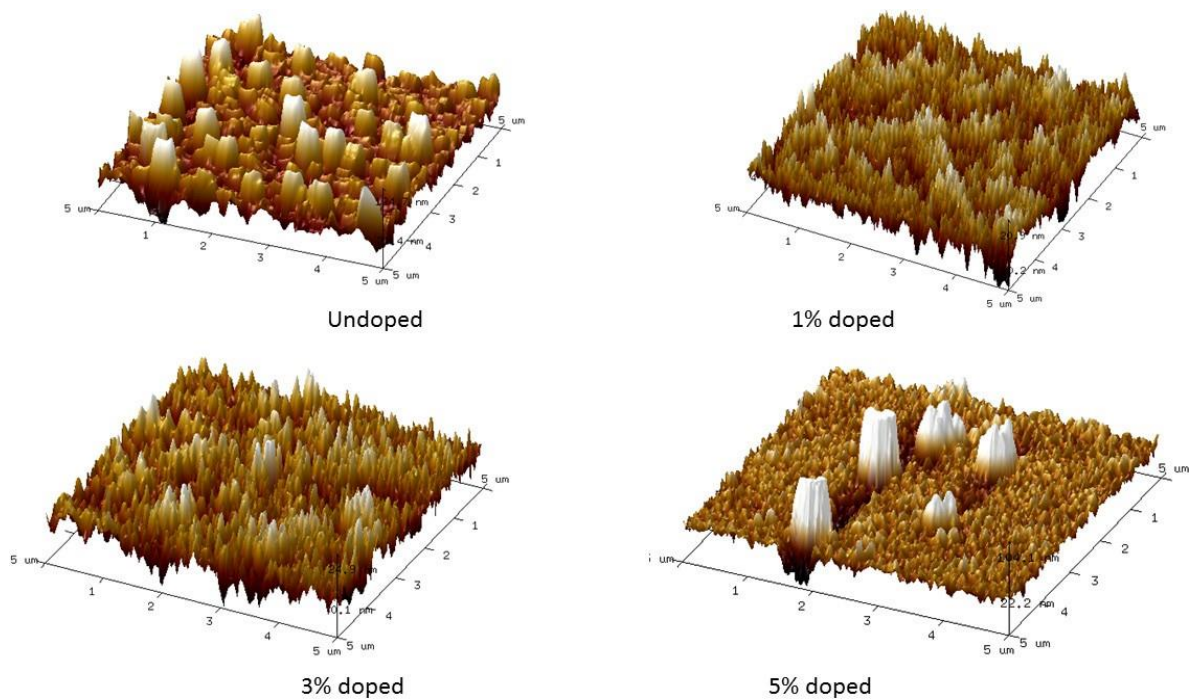

**Figure 6.** 3D AFM images of Cu:ZnO thin films.

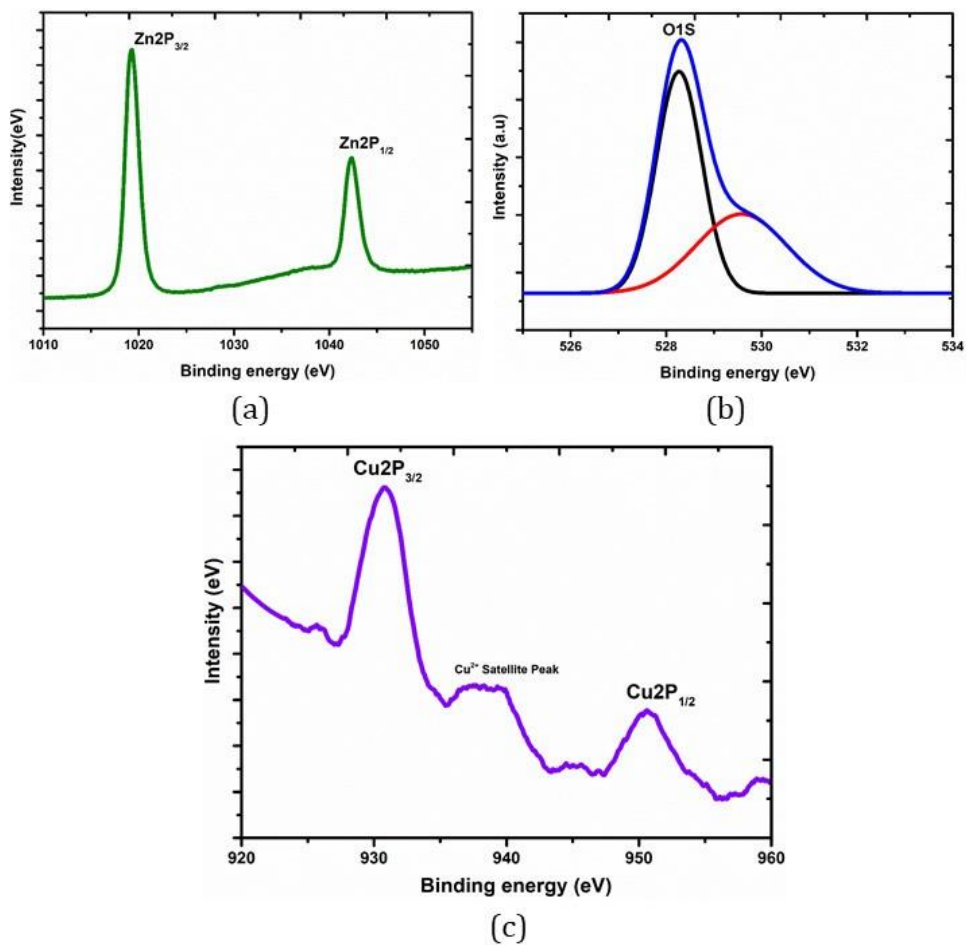

**Figure 7.** XPS spectra of (a) Zn2P (b) O1S (c) Cu2P.

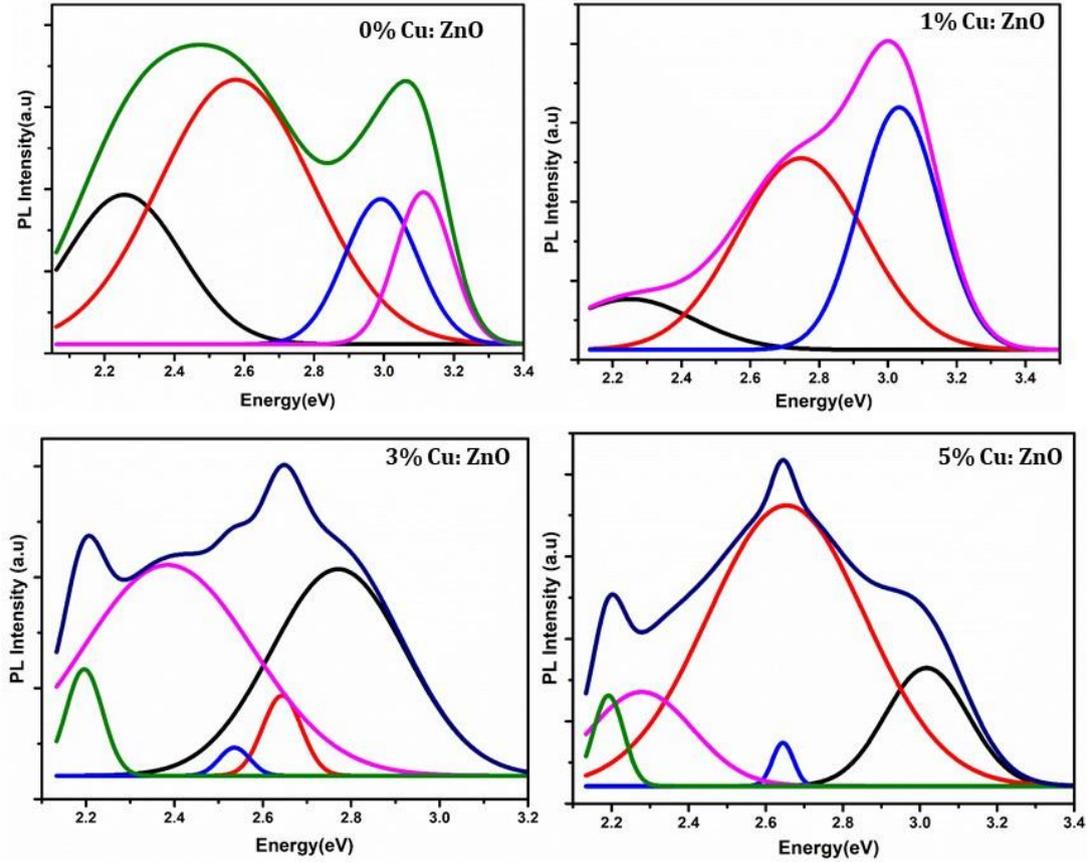

**Figure 8.** PL spectra of Cu:ZnO thin films (Gaussian fit).

The glancing angle XRD patterns for the Cu:ZnO thin films are shown in figure 3. These films show a polycrystalline nature assigned to hexagonal wurtzite ZnO structure (JCPDS 36-1451) with growth orientation along (0 0 2), (1 0 1), (1 0 2) and (1 0 3). The most intense diffraction peak is observed at (1 0 1) orientation for undoped ZnO thin films, which is changed to (0 0 2) orientation upon doping. The angle shift is due to the compression and relaxation of the host ZnO lattice upon Cu doping [26]. The presence of vacancies and interstitial defects is the principal origin of this lattice distortion. The absence of growth phases other than ZnO phases within the x-ray detection limit confirms the effective doping of copper into the ZnO lattice. The variations in crystalline size, lattice strain and dislocation density of the films upon doping are listed in table 1. Figure 4 depicts the variation in the crystalline size and lattice strain on copper concentration in the films. The enhancement in the crystalline size and decrement in the lattice strain confirms the crystallization of the ZnO nanostructure upon Cu doping.

Figures 5 and 6 show the 2D and 3D AFM images for the Cu:ZnO thin films. The surface topology of the films was determined using Bruker NanoScope analysis software. The AFM images depict homogenous, crack free and well-covered uniform surface morphology with pea-shaped grains. The probed films are composed of several grains, which confirms the polycrystalline nature of the films. The grain size reduced with the incorporation of Cu for 1% and 3% doping concentration, but increased at 5% doping. The variations in the grain size can be attributed to the lattice strain arising in the films due to the difference in the ionic radii of the Cu and Zn atom [27]. The increase in the number of nucleation centers during the Cu incorporation also contributes to the reduction in the grain size. The average surface roughness of 0%, 1%, 3% and 5% Cu-doped ZnO thin films is 26.9, 5.96, 7.00 and 16.6 nm. The smooth surface formed due to the Cu doping results in the reduction of light scattering, which enhances the optical and nonlinear optical properties of the film [27].

**Table 2.** Peak positions and proposed origin of defects.

| Peak position (eV) | Proposed origin | Reference |
| --- | --- | --- |
| 3.12 | NBE | [32] |
| 3.02, 3.00, 2.99 | Interstial zinc vacancy ($Zn_i$) and Cu acceptor levels | [33–36] |
| 2.80, 2.76 | Ionized zinc interstial ($Zn_i+$) and Cu acceptor levels | [33, 28, 40] |
| 2.66, 2.65, 2.64 | Oxygen anyisite defect site ($O_{zn}$) and $Cu^{2+}$ and $Cu^{+}$ acceptor levels | [33–39, 28] |
| 2.50, 2.45 | Cu impurities | [28, 40] |
| 2.40, 2.33 | Oxygen vacancy defect ($V_o$) | [33–35] |
| 2.20, 2.19 | Interstial oxygen ($O_i$) and hydroxyl group (OH) | [33] |

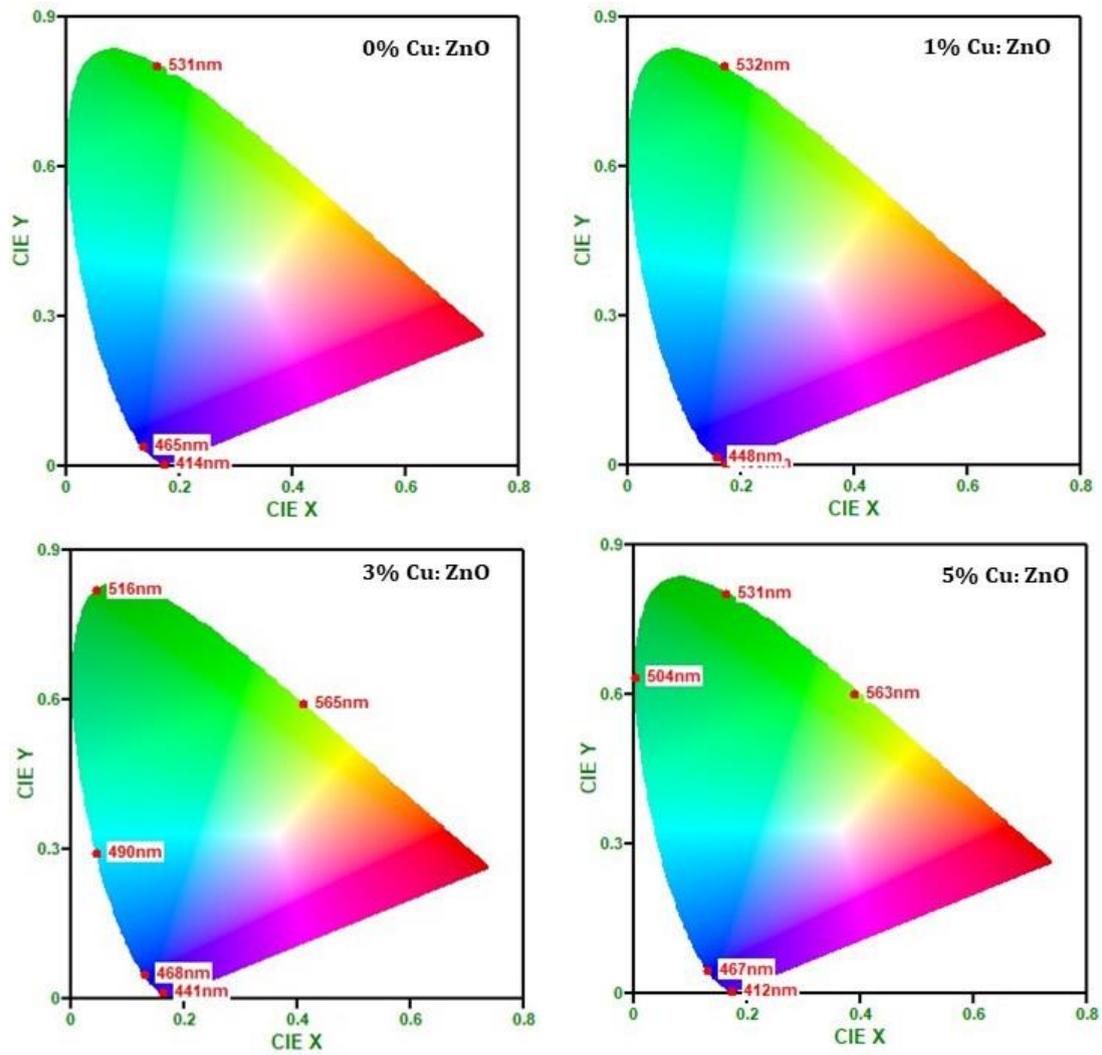

**Figure 9.** CIE diagram of Cu:ZnO thin films.

### 3.2 XPS studies

The surface properties and oxidation state of the deposited films were investigated using XPS analysis. Figure 7 depicts the XPS spectra of 3% Cu:ZnO thin films. The binding energies of Zn, O and Cu were calibrated with respect to the C1S peak at 286 eV as the reference peak. The core-level spectra of Zn2P in figure 7(a) show that the two peaks situated at 1019.4 and 1042.5 eV are attributed to $Zn2P_{3/2}$ and $Zn2P_{1/2}$. The energy difference between the two core-level Zn2P peaks is 23.1 eV, which is in line with earlier reports on ZnO [28-30]. This observation also confirms the complete oxidation of Zn in the Cu:ZnO thin films and the +2 valence state of the atom. It is observed from figure 7(b) that the core-level spectra of the O1S is not geometrically symmetric and the presence of multiple overlapping components can be observed from the spectra. These overlapping peaks can be identified by implementing a Gaussian peak fitting method [28]. The fitting results in two peaks for the O1S spectra centered at 528.27 and 529.54 eV. The peak at 528.27 eV is due to the presence of lattice oxygen in the form of Zn-O binding and the peak at 529.54 is attributed to the presence of oxygen vacancy defects in the films [28]. The XPS spectra of Cu2P in figure 7(b) show two signature peaks of the Cu element centered at 930.7 and 950.6 eV, which correspond to $Cu2P_{3/2}$ and $Cu2P_{1/2}$ [29]. The peaks confirm the successful doping of the Cu atom into the ZnO lattice. The presence of the well-known satellite peak observed for Cu2p spectra at 939.9 eV is an indication of the $Cu^{2+}$ oxidation state of the atom in the investigated films. The shift observed in the Cu2P peaks from the reported values [29–31] is possibly due to the creation of defect states in the films upon doping

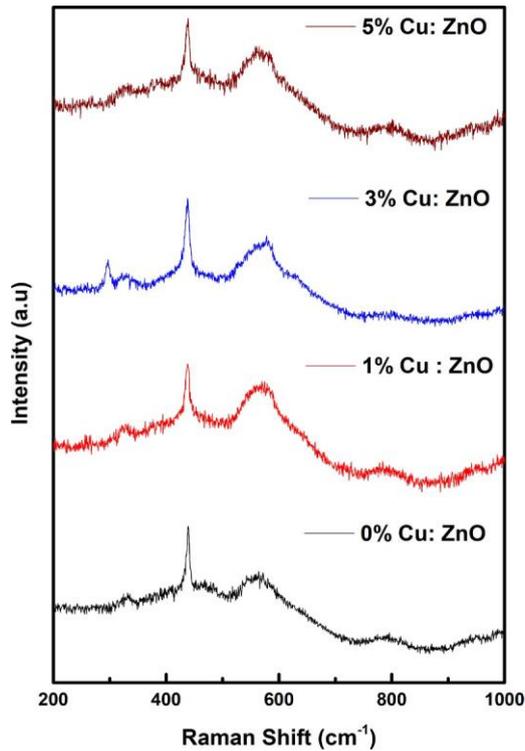

**Figure 10.** Raman spectra of Cu:ZnO thin films.

### 3.3 PL and Raman spectroscopy studies

Room-temperature PL spectra of pure and 1%, 3% and 5% Cu-doped ZnO nanostructures at an excitation wavelength of 325 nm consisting of near band edge (NBE) emission and a wide visible emission band are shown in figure 8. Gaussian fitting is performed on the spectra in order to determine the locus of the luminescent centers and to explore the origin of defects in the nanostructures. The Gaussian fit of undoped ZnO is dominated by four different emissions centered at 398, 414, 465 and 531 nm, corresponding to NBE UV emission, violet, blue and green emission. The NBE emission is due to the transition of electrons from the conduction band to the singly-ionized oxygen vacancy defect (Vo+) located at ~0.18 eV above the valence band [32]. The violet emission peak at 2.99 eV is attributed to the transition of electrons from the zinc interstial (Zni) defects to the valence band. The broad blue emission peak centered at 2.66 eV is due to the possible transition of electrons from the conduction band to the antisite oxygen site (Ozn), which act as a deep-level acceptor defect [32–37]. The spectrally broad PL peak centered at 2.33 eV is attributed to green luminescence, which is believed to stem from the surface of the ZnO thin films. In the present case, the emission is attributed to the recombination of electrons in the conduction band to holes trapped in the deep-level oxygen vacancy (Vo) [32, 37].

The emission behavior of the Cu ion is sensitive to the doping concentration and defects already present in the host ZnO lattice. The incorporation of Cu at 1%, 3% and 5% has speckled the PL spectra. Cu doping to the ZnO lattice results in the shifting of UV emission observed at 3.12 eV for undoped ZnO to the visible region. This peculiar behavior observed is due to the increase in non-radiative recombination mechanisms in the ZnO lattice [28]. The spectra also show a notable enhancement in the deep-level emission observed at the violet, blue and green emission centers. The incorporation of Cu ions into the ZnO lattice induces more defect states between the valence and conduction bands, causing a bandgap narrowing effect, as evidenced in figure 2. The NBE violet emission observed at 3.02 and 3.00 eV for 1% and 5% Cu:ZnO is due to the transition of electrons from the conduction band to the intrinsic Zni defect site and Cu acceptor levels [33]. Violet-blue emission centered at 2.76 and 2.80 eV for 1% and 3% Cu:ZnO films was attributed to the presence of the ionized Zni defect site and Cu acceptor levels below the conduction band, which traps electrons and results in the transition to the valence band [33–35]. The appearance of blue emission at 2.64 and 2.65 eV for 3% and 5% Cu:ZnO corresponds to the transition of electrons from the $Cu^{2+}$ donor level to $Cu^{+}$ acceptor level [28]. The absence of blue emission in undoped ZnO confirms that Cu ions can act as a luminescent activator, which can effectively regulate the luminescence behavior of ZnO [28, 39]. The blue-green luminescent center observed at 2.50 and 2.45 eV for 3% and 5% Cu:ZnO nanostructures was attributed to the oxygen vacancy defects and Cu impurities present in the films. The green PL emission observed at 2.32, 2.33 and 2.40 eV for 1%, 3% and 5% Cu:ZnO thin films appeared due to the presence of oxygen vacancy defects, which traps the photo-generated holes and results in the recombination with the electron trapped in the shallow donor level. Upon Cu doping, the broadness of the green luminescent peak increases due to the increase in the density of defect states, which results in green PL emission. The yellow emission peaks observed for 3% and 5% doped films at 2.19 and 2.20 eV were assigned to the interstial oxygen (Oi) and absorbed hydroxyl group (OH) [32, 36–38]. Table 2 depicts the wavelength of the PL emission observed and proposed origin of its defects. A notable enhancement in the violet, blue and green luminescence was observed, which infers the enlarged defect states in the samples upon Cu doping to ZnO nanostructures.

The obtained PL spectra of the Cu:ZnO thin films were transposed to the Commission Internationale de l'Eclairage (CIE 1931) diagram in order to study the photometric characteristics [40]. The chromatic coordinates ($x,y$) for each emission were calculated [41] and are represented in figure 9. The changes in the ($x, y$) coordinates of the green emission and the formation of new color centers at (0.045, 0.294) for 3% Cu:ZnO and (0.003, 0.633) for 5% Cu:ZnO thin films shows the effect of Cu doping in the photometric properties of the ZnO thin films. The variation observed in the CIE coordinates also endorses the active contribution of Zn, Zni, Oi, Cu and $Cu^{2+}$ defect states in the fluorescent properties of the Cu:ZnO thin films.

Figure 10 shows the Raman spectra of the Cu:ZnO thin films. The films are probed by a diode-pumped solid state laser of 532 nm wavelength and at an acquisition time of 20 s. The wurtzite ZnO belongs to the P63mc-C6v4 space group with four atoms per the unit cell. The group theory analysis represents the relation of the vibrational modes of ZnO at the center point of Brillouin zones as 2A1 + 2E1 + 2B1 + 2E2 [42]. The Raman spectra reveal peaks at 333, 439 and 558 cm$^{-1}$, corresponding to E2H-E2L, E2H and A1 (LO) phonon modes, respectively. In the measurement range of 200-1000 cm$^{-1}$, the peak at about 439 cm$^{-1}$ dominates the Raman spectra of all the Cu:ZnO nanostructures. This peak is assigned to the E2H mode, which is attributed to the vibration of the oxygen atom in the lattice and is a characteristic feature of wurtzite ZnO with high crystalline behavior [33, 39]. The presence of a signature peak at 439 cm$^{-1}$ confirms the XRD results obtained for the films. The peaks observed at 333, 325 and 332 cm$^{-1}$ for 0%, 1%, 3% and 5% Cu:ZnO films corresponds to the E2H-E2L mode associated with crystal disturbances that arise due to the vibration of the oxygen atom and zinc sublattice. The broad peaks observed at 558, 570, 571 and 569 cm$^{-1}$ for the films were assigned to the A1(LO) mode, which appears only when the *C*-axis of the wurtzite ZnO remains parallel to the incident light [33]. The broadening of this Raman peak confirms the presence of Zni and oxygen defects and its enhancement upon Cu incorporation. In 3% Cu:ZnO films a Cu related Raman peak is observed around 297 cm$^{-1}$ assigned to the A1g phonon mode, which is attributed to the vibration of the oxygen atoms in the ZnO lattice. Table 3 shows the assignment of Raman peaks and phonon modes of the Cu:ZnO nanostructures at different Cu concentration. It is observed that with increasing Cu concentration a clear shift in the peak position towards a higher and lower wavenumber side. The probable mechanisms behind the peak shifting in the Raman spectra refer to phonon localization by defects, spatial confinement within the nanostructure boundaries and laser self-heating effects [27].

**Table 3.** Raman peaks and phonon modes of Cu:ZnO thin films.

| 0% F:ZnO | | | 1% F:ZnO | | | 3% F:ZnO | | | 5% F:ZnO | | |
|---|---|---|---|---|---|---|---|---|---|---|---|
| Raman peak (cm$^{-1}$) | Phonon mode | Shift (cm$^{-1}$) | Raman peak (cm$^{-1}$) | Phonon mode | Shift (cm$^{-1}$) | Raman peak (cm$^{-1}$) | Phonon mode | Shift (cm$^{-1}$) | Raman peak (cm$^{-1}$) | Phonon mode | Shift (cm$^{-1}$) |
| 333 | E2H-E2L | — | 333 | E2H-E2L | — | 325 | E2H-E2L | −8 | 332 | E2H-E2L | −1 |
| 439 | E2H | — | 439 | E2H | −2 | 437 | E2H | −2 | 437 | E2H | −2 |
| 558 | A1(LO) | — | 570 | A1(LO) | +12 | 571 | A1(LO) | +13 | 569 | A1(LO) | +11 |
|  |  |  |  |  |  | 297 | A1g | — |  |  |  |

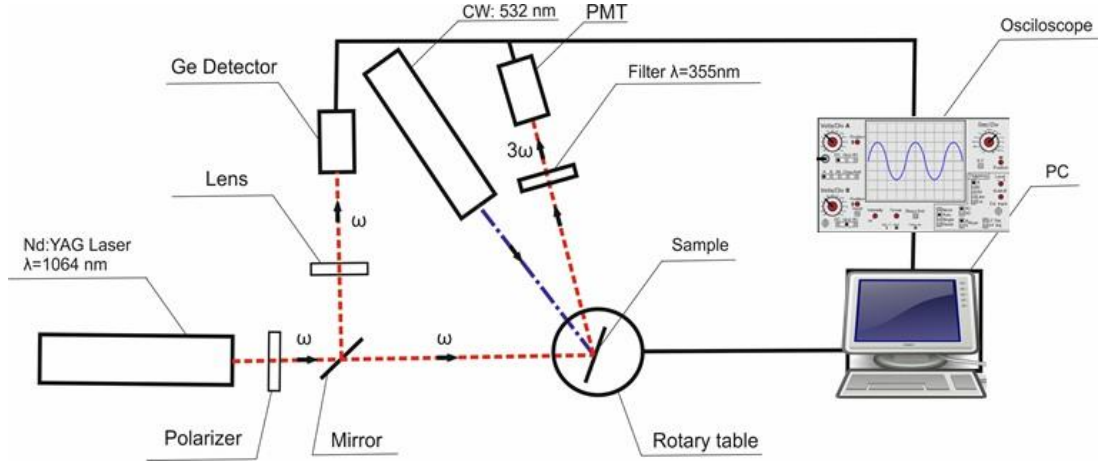

**Figure 11.** Experimental setup of THG measurement.

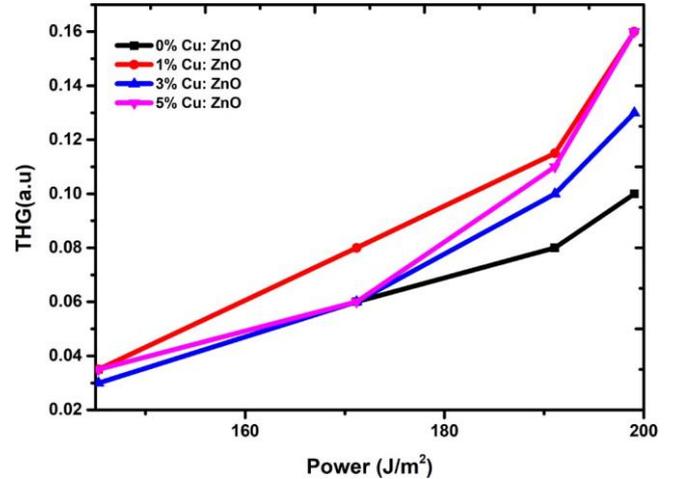

**Figure 12.** THG signal versus input laser power for Cu:ZnO thin films.

### 3.4. Third harmonic process

*3.4.1. THG.* In the THG technique, a high-intensity laser beam at the frequency $\omega$ interacts with a nonlinear medium and generates an additional beam at a frequency of $3\omega$. This is considered one of the most powerful techniques for accurately assessing the ultra-fast electronic nonlinear response of the material probed [43]. The contributions from thermal, orientation and vibrational effects can be excluded by employing this technique. Figure 11 shows the experimental setup used for the THG measurement and figure 12 corresponds to the THG signal response of the Cu:ZnO thin films at different Cu doping concentration. A Nd:YAG laser at 1064 nm wavelength with 8 ns pulse rate and 10 Hz repetition rate was used as the fundamental beam. The effect of photo induction with a coherent CW laser light with a wavelength of 532 nm on the level of the THG signal was examined. The input power of the fundamental laser beam was varied from 140-200 J m$^{-2}$ using a Glan polarizer. It is clear from figure 10 that the THG signal intensity has been enhanced upon increasing the Cu doping concentration and input laser power. The highest THG conversion efficiency was exhibited by 1% Cu:ZnO films followed by 5% and 3% Cu-doped ZnO films. The defect-assisted enhancement in the photoexcitation and relaxation process that occurred in the probed Cu:ZnO nanostructures can be the possible reason for the increased $3\omega$ efficiency [44].

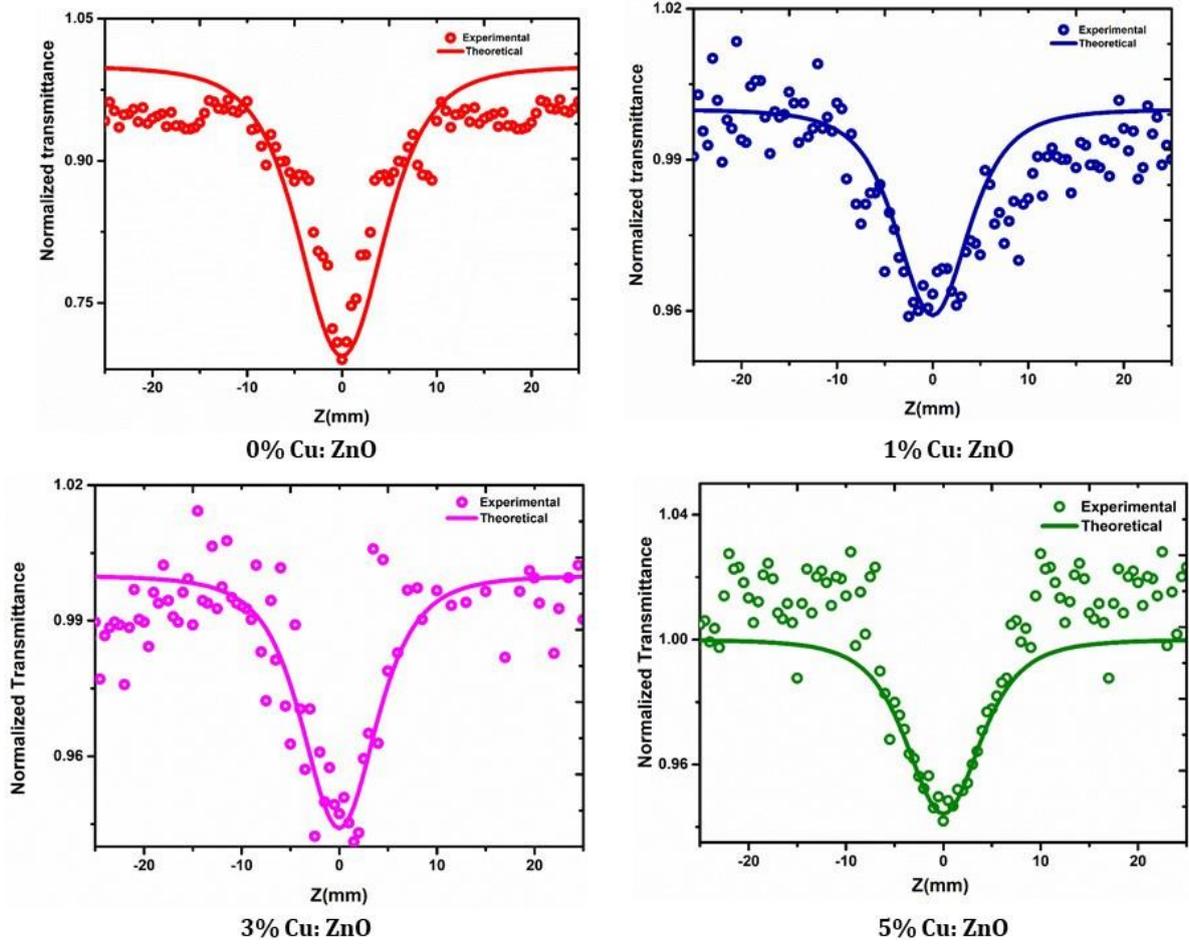

**Figure 13.** OA Z-scan trace.

*3.4.2. Z-scan technique.* The intensity-dependent nonlinear absorption ($\beta$) and nonlinear refractive index ($n_2$) process for the Cu:ZnO thin films at different doping concentration were studied by single-beam Z-scan technique [45, 46] in continuous laser regime at an input intensity of 9.46 × 10$^6$ W m$^{-2}$. The open aperture (OA) Z-scan measurements were performed to determine the $\beta_{eff}$ of the films. The intensity dependent $\beta_{eff}$ is given by

$$\alpha_0 = \alpha + \beta_{eff} I, \qquad (1)$$

where $I$ is the input laser intensity and $\alpha$ is the linear absorption coefficient of the investigated film. The OA Z-scan traces of the Cu:ZnO thin films at different concentration are shown in figure 13. All films exhibit well-defined normalized valley at the focus, which infers the reverse saturable absorption process (RSA). The signature of the RSA process in the films can be attributed to any of the phenomena such as excited state absorption, two photon absorption (TPA), free carrier absorption (FCA), induced nonlinear scattering or a combination of any of these processes. It is noteworthy to consider the energetics required for the TPA process to occur. TPA is allowed only when the incident energy of the light source is less than the energy band gap $E_g$, but greater than $E_g/2$ of the investigated film [47]. In the Cu:ZnO thin films, the incident laser source energy is equal to 1.95 eV, thus satisfying the above necessary condition. If the origin of nonlinear absorption is due to the TPA process alone then the $\beta_{eff}$ will be a constant independent of intensity. But in the investigated films, $\beta_{eff}$ decreases with an increase in intensity. This behavior observed in the films can be attributed to the sequential TPA process reported by Abdlefdil *et al* [51]. Therefore, the origin of the nonlinear absorption in the probed films is due to the FCA-induced TPA process.

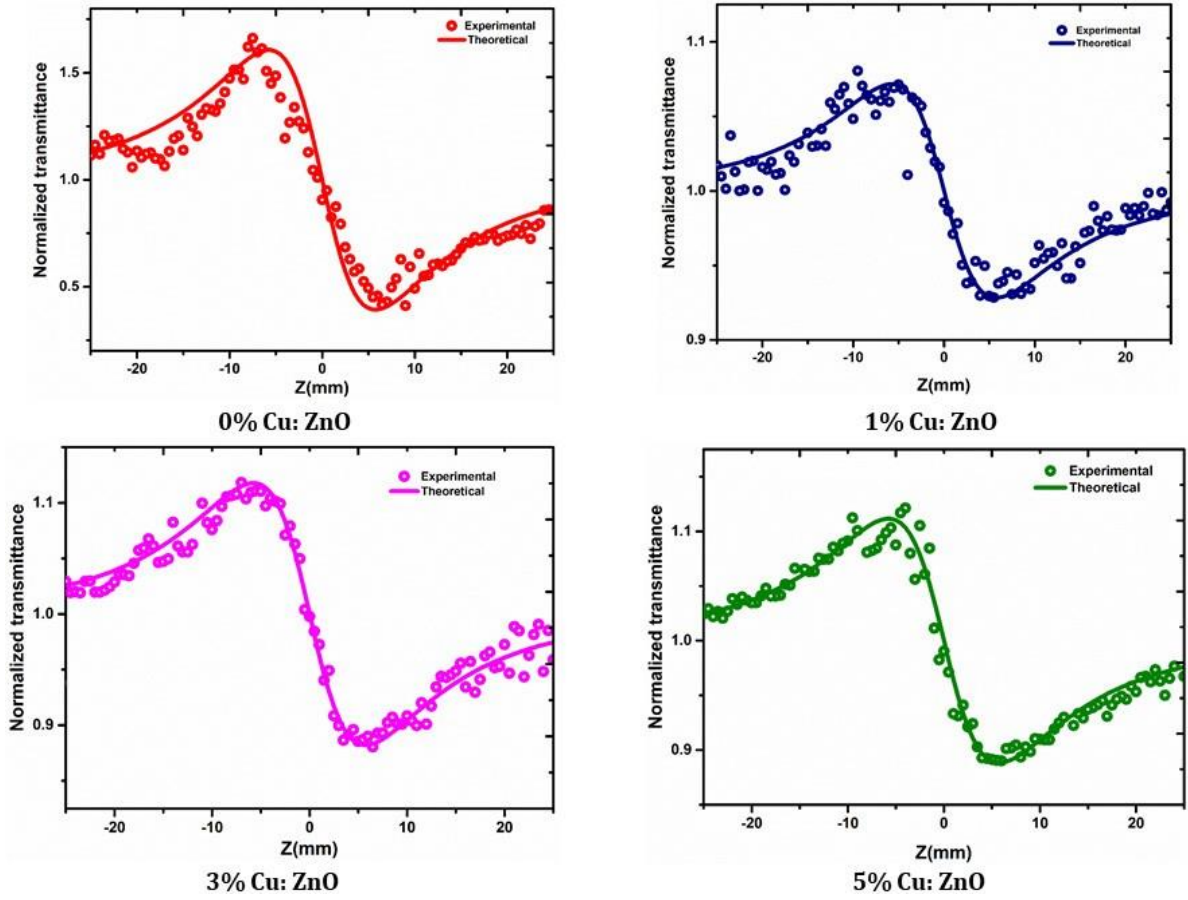

**Figure 14.** CA Z-scan trace.

**Table 4.** Third-order nonlinear optical parameters of Cu:ZnO thin films.

| Conc. (%) | $\beta_{eff} \times 10^{-2}$ (cm W$^{-1}$) | $n_2$ (esu) | $\chi_R^{(3)} \times 10^{-3}$ esu | $\chi_I^{(3)} \times 10^{-4}$ esu | $x^{(3)} \times 10^{-3}$ esu |
|---|---|---|---|---|---|
| 0 | 1.05 | −0.032 | −0.35 | 0.199 | 0.35 |
| 1 | 3.67 | −0.081 | −1.75 | 1.94 | 1.76 |
| 3 | 4.28 | −0.127 | −2.76 | 2.25 | 2.77 |
| 5 | 5.04 | −0.111 | −2.42 | 2.66 | 2.43 |

The closed aperture (CA) Z-scan measurements were employed to determine the nonlinear refractive index of the films. Figure 14 shows the CA trace of the Cu:ZnO thin films. In the CW region, the nonlinear refraction occurs in the films via a non-radiative process such as thermal, electrostrictive or mechanical, etc [48, 49]. The peak valley difference $\Delta T_{Pv}$ ~ 1.7 times the Rayleigh range ($Z_R$) of the laser confirms that the refractive nonlinearity exhibited by the Cu:ZnO nanostructures is thermal in nature. The thermal nonlinearity exhibited will produce a thermal lensing effect where all the films act as a negative lens with self-defocusing behavior [50]. Table 4 shows the third-order nonlinear optical parameters of the Cu:ZnO thin films. The nonlinear optical coefficients were calculated by fitting the experimental Z-scan curve by nonlinear transmission equations, as given by Bahea *et al*. The third-order nonlinear optical susceptibility ($x^{(3)}$) and its increment upon Cu doping concentration is caused by changes in the nonlinear absorption coefficient due to enhancement in the electronic transitions due to the different intermediate states created in the films upon doping and variations in the nonlinear refractive index through the local heating effects [49-51]. In other words, the XPS, PL, XRD and Raman spectra results prove the presence of defects in the Cu:ZnO nanostructures. A small amount of energy is absorbed by the particles as well as the defect states during irradiation of the laser. When the Cu concentration and defects are high, the thermal excitement of the particle due to the localized heating effect will result in the change of $n_2$ and electronic transition will be enhanced due to the formation of intermediate defect states, which results in the change of $\beta_{eff}$. So, the enhancement in nonlinearity of Cu-doped ZnO nanostructures is revealed.

## 4. Conclusions

ZnO and Cu:ZnO thin films with 1%, 3% and 5% doping concentration were grown by spray pyrolysis technique. The surface morphology of the films shows uniform pea-shaped grains upon Cu incorporation into the ZnO lattice. The shift in core-level peak position of Zn2P and Cu2P is attributed to the presence of defect states in the films. The PL spectra show increased emission intensity along the violet, blue and green color centers on Cu-doped ZnO thin films with a blue emission peak observed for 5% Cu:ZnO as the most intense among them. The shift in the UV emission peak observed around 396 nm for undoped ZnO to the visible region suggests the increased non-radiative process in the films on Cu doping. The XRD analysis of the films shows an enhancement in the crystalline size with a preferred growth direction (0 0 2) parallel to the $C$-axis. The phonon modes observed in Raman spectra confirm the structural results proposed by XRD analysis. The observed blueshift and redshift in the Raman modes are attributed to defect states formed in the films. The single-beam Z-scan measurements show a one-order increase of $x(3)$ from $10^{-4}$ to $10^{-3}$ esu. The THG studies conducted on the films show an increase in the THG signal intensity and the possibility of using this material in frequency tripling applications.

## Acknowledgments

A portion of this research was performed using facilities at CeNSE, funded by the Ministry of Electronics and Information Technology (MeitY), Government of India, and located at the Indian Institute of Science, Bengaluru.

For K.O, J.J., A.W., P.R., K.I.V this work is a part of a project that has received funding from the European Union's Horizon 2020 research and innovation program under the Marie Skłodowska-Curie grant agreement No. 778156.